\documentclass[twocolumn,showpacs,preprintnumbers,amsmath,amssymb]{revtex4}

\usepackage{graphicx}
\usepackage{dcolumn}
\usepackage{bm}

\begin{document}

\newcommand{\RR}[1]{[#1]}
\newcommand{\intsum}{\sum \kern -15pt \int}
\newfont{\Yfont}{cmti10 scaled 2074}
\newcommand{\Y}{\hbox{{\Yfont y}\phantom.}}
\def\O{{\cal O}}
\newcommand{\bra}[1]{\left< #1 \right| }
\newcommand{\braa}[1]{\left. \left< #1 \right| \right| }
\def\Bra#1#2{{\mbox{\vphantom{$\left< #2 \right|$}}}_{#1}
\kern -2.5pt \left< #2 \right| }
\def\Braa#1#2{{\mbox{\vphantom{$\left< #2 \right|$}}}_{#1}
\kern -2.5pt \left. \left< #2 \right| \right| }
\newcommand{\ket}[1]{\left| #1 \right> }
\newcommand{\kett}[1]{\left| \left| #1 \right> \right.}
\newcommand{\scal}[2]{\left< #1 \left| \mbox{\vphantom{$\left< #1 #2 \right|$}}
\right. #2 \right> }
\def\Scal#1#2#3{{\mbox{\vphantom{$\left<#2#3\right|$}}}_{#1}
{\left< #2 \left| \mbox{\vphantom{$\left<#2#3\right|$}}
\right. #3 \right> }}

\title{
Selectivity of the Nucleon Induced Deuteron Breakup and 
Relativistic Effects}

\author{H.\ Wita{\l}a}
\author{J.\ Golak}
\author{R.\ Skibi\'nski}
\affiliation{Institute of Physics, Jagiellonian University,
PL-30059 Cracow, Poland}
\date{\today}

\begin{abstract}
 Theoretical predictions for the nucleon induced 
deuteron breakup process based on solutions of 
 the  three-nucleon  Faddeev
 equation including  such relativistic features as the 
relativistic kinematics and 
 boost effects are presented.  Large changes of the breakup cross section in 
some complete configurations are found at higher energies. The 
predicted relativistic effects, which are mostly of 
dynamical origin, seem to be supported by existing data.
\end{abstract}

\maketitle


Recent studies of elastic nucleon-deuteron (Nd) scattering
and nucleon induced deuteron breakup
 revealed a number of cases where the nonrelativistic description
based only on pairwise nucleon-nucleon (NN) 
forces  is insufficient to explain the three-nucleon (3N) data. This happens 
in spite of the fact, that these high precision   NN potentials 
 describe very well the
NN data set  to about 350 MeV laboratory energy.
 Those findings  extended  exploration of  the properties
of three-nucleon forces (3NFs)  to the reactions in the 3N continuum. 
Such  forces appear for the first time in the 3N system
where they provide an additional contribution to a predominantly
pairwise potential energy of three nucleons. 
Generally speaking, the studied
discrepancies between a theory based only on NN potentials  and experiment
  become larger  for increasing energy of the 3N system. Adding
 a 3N force  to the  pairwise interactions  leads 
in some cases
to a better description of the data. 
 The best studied example is the discrepancy  for 
 the elastic 
angular distribution in the region of its minimum and at backward
angles~\cite{wit98,wit01,sek02}. 
This clear discrepancy  can be removed at energies below  $\approx 100$~MeV by
adding modern 3NFs to the nuclear Hamiltonian. Such a 3NF, 
mostly of the $2\pi$-exchange character,  must be
adjusted individually with each  NN potential  to
the experimental binding energy of $^3$H and
$^3$He. 
 At energies higher than  $\approx 100$~MeV current
3NFs  improve only partially the
description of cross section data and the remaining
 discrepancies, which  increase with energy,
 indicate the  possibility of relativistic effects.
The need for a  relativistic description of 3N scattering was also
raised  when precise
measurements of the total cross section for neutron-deuteron (nd)
interaction~\cite{abf98} were analyzed within the framework of
 the nonrelativistic Faddeev calculations~\cite{wit99}.
Also there the NN forces alone were
 insufficient to describe the data above  $\approx 100$~MeV.
 The effects due to relativistic kinematics considered in
 ref.~\cite{wit99}
were comparable at higher energies to the effects due to 3NFs.
 This demonstrates the
importance of a study taking  relativistic
effects in the 3N continuum into account.

Recently first results of relativistic 3N Faddeev calculations for 
the elastic Nd scattering have become available~\cite{wit2005}. 
The relativistic 
formulation applied was of the instant form of relativistic 
dynamics~\cite{keister91}. 
A starting point of this formulation for 3N scattering is the Lorentz boosted 
NN potential  $V( \vec k, \vec k~' ;  \vec P~)$ which generates 
the two-nucleon (2N) boosted t matrix $t( \vec k, \vec k~' ;  \vec P~)$  
in a moving frame via a standard relativistic 2N Lippmann-Schwinger 
equation
\begin{eqnarray}
 &&t( \vec k, \vec k~' ;  \vec P~) = 
 V( \vec k, \vec k~' ;  \vec P~)  +  
\int d^3k'' \cr 
&&{ 
{V( \vec k, \vec k~'' ;  \vec P~)
 t( \vec k~'', \vec k~' ;  \vec P~) } 
\over{ \sqrt{ {( 2\omega( {\vec k}^{\ '}) )^{\ 2} + {\vec P}^{\ 2}} }  
-  \sqrt{ {(2\omega(  {\vec k}^{\ ''}) )^{\ 2} 
+ {\vec P}^{\ 2}} } + i\epsilon  } } .
\label{eq2a}
\end{eqnarray}
The NN potential in an arbitrary moving frame $V(\vec P \, )$ is obtained from 
the interaction $v$ defined in the two-nucleon c.m. system by~\cite{relform1}
\begin{eqnarray}
V(\vec P \, ) \equiv \sqrt{( 2\omega(\vec k) + v)^2 + {\vec P}^{\ 2}} 
 - \sqrt{ (2\omega(\vec k))^2 + {\vec P}^{\ 2}}.
\label{eq1}
\end{eqnarray}
The relativistic kinetic energy of three equal 
mass (m) nucleons  in their c.m. system can
 be expressed  by the relative momentum  $\vec k$ 
 in one of the two-body subsystems and momentum of the third nucleon  
 $\vec q$ (the total momentum of the two-body subsystem is 
then $\vec P = -\vec q$) as
\begin{eqnarray}
H_0 &=& \sqrt{(2\omega({\vec k}))^2 + {\vec q}^{\ 2}} 
+ \sqrt{m^2 + {\vec q}^{\ 2}} ,
\label{eq1an1}
\end{eqnarray}
where $2\omega({\vec k}) \equiv 2\sqrt{m^2 + {\vec k}^{\ 2}}$ is 
the momentum dependent 
2N mass operator. 

The Nd scattering with neutrons and protons interacting
through a NN potential $V$ alone  is described in terms of a breakup operator T
satisfying the Faddeev-type integral equation~\cite{wit88,glo96}
\begin{eqnarray}
T\vert \phi >  &=& t P \vert \phi > + t P G_0 T \vert \phi > .
\label{eq1a}
\end{eqnarray}
The permutation operator  $P=P_{12}P_{23} + P_{13}P_{23}$ is given in terms
of the transpositions $P_{ij}$, which interchange nucleons i and j. 
The incoming state  
$ \vert \phi > \equiv 
\vert \vec q_0 > \vert \phi_d > $ describes the free
nucleon-deuteron motion with the relative momentum $\vec q_0$ and
the deuteron wave function $\vert \phi_d >$. 
The $G_0 \equiv {{1}\over{E + i\epsilon - H_0}}$ is the free 3N propagator 
 with  total 3N c.m. 
energy $E$ expressed in terms of the initial
neutron momentum $\vec q_0$ relative to the deuteron
\begin{eqnarray}
E &=& \sqrt{(M_d)^2 + {\vec q}_0^{\ 2} } + \sqrt{m^2 + {\vec q}_0^{\ 2}} ,
\label{eq2d}
\end{eqnarray}
where $M_d$ is the deuteron rest mass. 

The  transition operators for elastic  scattering, U, and 
 breakup, $U_0$, are  
  given in terms of T by~\cite{wit88,glo96}
\begin{eqnarray}
U  &=& P G_0^{-1} + P T \cr
U_0  &=& (1 + P) T .
\label{eq1c}
\end{eqnarray}
The state $ U_0 \vert \phi >$ is projected onto
the state $\vert {\phi}_0 >$ which describes
the free motion of the three outgoing nucleons in the 3N c.m. system in terms
of the relative momentum of the 2N subsystem $\vec k_{3N~c.m.}(2-3)$, defined 
in the 3N c.m.,    and momentum of the spectator nucleon 
 $\vec q$ defined above:
 $\vert {\phi}_0 > \equiv \vert \vec k_{3N~c.m.}(2-3)~ \vec q >_1$.
 This leads to the breakup transition amplitude
\begin{eqnarray}
< {\phi}_0 \vert U_0 \vert \phi > = \sum_i 
 {_i}< \vec k_{3N~c.m.}(j-k)~ \vec q_i \vert T \vert \phi >  .
\label{eqzz1}
\end{eqnarray}

The choice of the relative momentum $\vec k$ in the NN c.m. subsystem 
and the momentum $\vec q$ of the spectator nucleon in the 3N c.m. 
system to describe configuration of three nucleons is the most convenient 
in the relativistic case. In the nonrelativistic limit the momentum $\vec k$ 
reduces to the standard Jacobi momentum $\vec p$~\cite{glo96}.  
To solve  Eq.(\ref{eq1a})   numerically   partial wave decomposition is still 
required. The standard partial wave states  
$\vert p q \alpha > \equiv \vert p q (ls)j 
(\lambda{1\over{2}})IJ(t{1\over{2}})T >$~\cite{glo96}, however, are 
generalized in the relativistic case due to the choice of the NN-subsystem 
momentum $\vec k$ and the total spin s both defined in the NN c.m. system. 
This lead to Wigner spin rotations when boosting to the 3N c.m. 
system~\cite{keister91,wit2005}, resulting in a more complex form for the 
permutation matrix element~\cite{wit2005} than used in the nonrelativistic 
case~\cite{glo96}. A restricted relativistic calculation with $j < 2$ 
partial wave states showed that Wigner spin rotations have only negligible 
effects~\cite{wit2005}. Due to this we neglected the 
Wigner rotations completely in the present study. 
 To achieve converged results at  energies up to $\approx 250$~MeV 
 all partial 
wave states with total
angular momenta of the 2N subsystem  up to $j \le 5$ have to be 
used and  all total
angular momenta of the 3N system up to $J=25/2$ taken into account.
 This leads to a system
of up to 143 coupled integral equations in two continuous variables
for a given total angular
momentum J and total parity $\pi=(-)^{l+\lambda}$ of the 3N system.
For the details of our relativistic formulation and of the numerical 
performance in the relativistic and nonrelativistic 
 cases we refer to Ref.~\cite{wit2005,wit88,glo96}.

In the present study we applied as a dynamical input a
 relativistic interaction $v$ generated from the nonrelativistic 
NN potential CDBonn~\cite{cdb} according to the analytical prescription of 
ref.~\cite{kam98}. 
This analytical transformation allows to obtain an exactly on-shell 
 equivalent to the CDBonn relativistic potential $v$ which 
provides the corresponding 
relativistic t matrix. 
 The boosted potential was not 
treated in all its complexity 
as given in ref.~\cite{kam2002} but a 
restriction to the leading order term in a $P/\omega$ and 
$v/\omega$ expansion  was made
\begin{eqnarray}
&&V(\vec k, \vec k~'; \vec P~)  =  
v(\vec k, \vec k~')~ \cr 
&&\times[~ 1  
 -  
{{\vec P}^{\ 2} \over{8\sqrt{m^2 + {\vec k}^{\ 2}}  
\sqrt{m^2 +  ( {\vec k}^{\, \prime} )^2 }}} ~ ] .
\label{eq2apr}
\end{eqnarray}
The quality of such an approximation has been checked by calculating 
the deuteron wave function $\phi_d(\vec k)$ of the  deuteron 
moving with  momentum $\vec P$ for a number of values corresponding 
to incoming nucleon lab. energy $\le 250$~MeV. The resulting deuteron binding 
energies and the deuteron D-state probabilities for the deuteron in 
 motion are close to the values for the deuteron at rest.  
 
\begin{figure}
\includegraphics[scale=0.59]{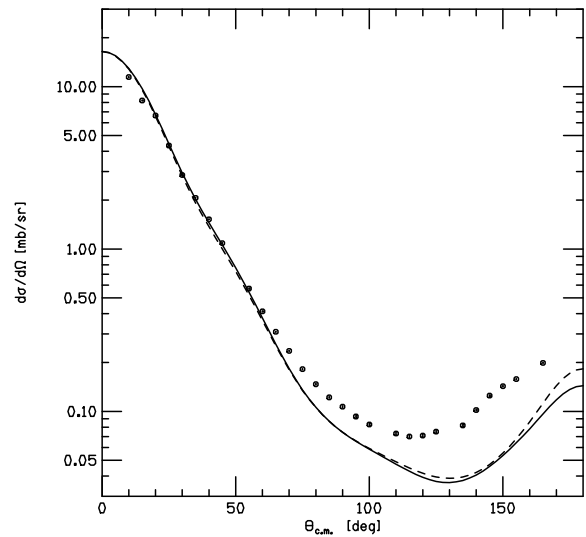}
\caption{
Differential cross section for elastic $Nd$ scattering at
250 MeV.
The dashed and solid lines are the results of
 the relativistic and nonrelativistic Faddeev calculations based on the
NN potential CDBonn. 
The  pd data are from \cite{hatanaka02}.
}
\label{fig1}
\end{figure}

 In Fig.~\ref{fig1} we show the nucleon angular distribution for 
 elastic nucleon-deuteron scattering at 
$E_{lab}^N =250$~MeV. It is 
seen that, like in a study of  Nd elastic scattering  in ref.~\cite{wit2005} 
 where the AV18~\cite{AV18} NN potential instead of CDBonn have been used, 
 relativistic effects 
for the cross section  are restricted 
only to the backward angles where relativity increases the nonrelativistic 
cross section. At other angles the effects are small. In spite of the 
fact that the relativistic phase-space factor increases with energy faster 
than the nonrelativistic one (at $250$~MeV their ratio amounts to 1.175), 
the relativistic nuclear matrix element 
outweighs this increase and leads for the cross section in a wide 
angular range to a relatively small relativistic effect.

The breakup reaction with three free outgoing nucleons in the final state 
 provides a unique possibility to access  
 the matrix elements  of the 
breakup operator T with specific values of momenta $\vert \vec k \vert$ and 
 $\vert \vec q \vert$ in a pointwise 
manner.  
Each exclusive  breakup configuration  
specified completely by 3N c.m. momenta $\vec k_i$ of outgoing  nucleons 
 requires three  matrix elements 
${_i}< \vec k(\vec k_j, \vec k_k), \vec q = \vec k_i \vert T  \vert \phi >$ 
 with $(i,j,k)=(1,2,3)$ and cyclical permutations, and $\vec k$ and $\vec q$ 
providing the total 3N c.m. energy
\begin{eqnarray}
E &=& \sqrt{4(m^2 + {\vec k}^2) + {\vec q}^{\ 2} } + \sqrt{m^2 + {\vec q}^{\ 2}} .
\label{eq2z}
\end{eqnarray}
This is entirely  different from the  elastic scattering where, due to  
continuum momentum distribution 
of nucleons inside the deuteron a broad range of $\vert \vec k \vert$- and 
$\vert \vec q \vert$-values 
 contributes   
to the elastic scattering transition matrix element. That particular 
selectivity of the breakup singles out this reaction as a tool 
 to look for localized effects which when averaged are difficult 
 to see in elastic scattering. 
 
\begin{figure}
\includegraphics[scale=0.59]{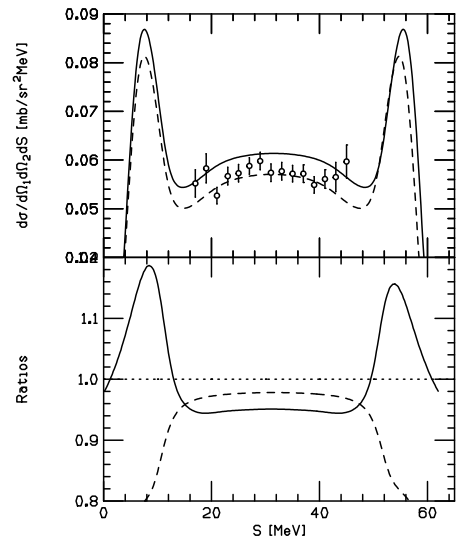}
\caption{
Five-fold differential cross section
 $d^5\sigma/d\Omega_1d\Omega_2dS$
 as a function of arc-length S
 for the  SSS  configuration (see text) of d(p,pp)n reaction
 at $65$~ MeV. The lab. angles of the outgoing protons are
$\theta_1 = \theta_2 = 54^{\circ}$ and $\phi_{12} = 120^{\circ}$.
The solid and dashed  lines are the results of
 the nonrelativistic and relativistic Faddeev calculations based on the
NN potential CDBonn.
The $65$~MeV pd data are from \cite{psi}. The lower part shows the ratios
of the relativistic and nonrelativistic phase-space factors (dashed line) and
 of the cross sections divided by the corrresponding phase-space
factor (solid line).
}
\label{fig2}
\end{figure}

\begin{figure}
\includegraphics[scale=0.59]{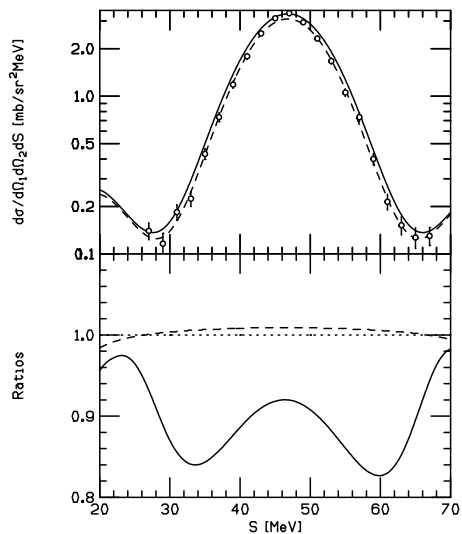}
\caption{
The same as in Fig.~\ref{fig2} for the QFS geometry (see text). The
 lab. angles of the outgoing protons are
$\theta_1 = \theta_2 = 44^{\circ}$ and $\phi_{12} = 180^{\circ}$.
The $65$~MeV pd data are from \cite{psiallet}.
}
\label{fig3}
\end{figure}

This selectivity of breakup helps to reveal 
 relativistic effects in the 3N continuum. Even at relatively low 
 incoming nucleon energy $E_{lab}^N = 65$~MeV 
 they can be clearly seen in cross sections of some exclusive 
  breakup configurations  as 
examplified in Figs.~\ref{fig2} and \ref{fig3}. 
 For the configuration of Fig.~\ref{fig2} the angles of the two outgoing 
 protons detected in coincidence  were chosen in such 
a way that for the arc-length $S \approx 30$~MeV  all 
three nucleons have equal momenta which  in the 3N c.m. system 
lie in the plane perpendicular to the 
beam direction (symmetrical space star (SSS) condition). For the configuration 
from Fig.~\ref{fig3} at the value of $S \approx 46$~MeV 
 the third, not observed nucleon is at rest 
in lab. system (quasi-free scattering (QFS) geometry). In these two 
 breakup configurations the inclusion of relativity lowers the cross section: 
 by $\approx 8 \%$ in the 
case of SSS and by $\approx 10 \%$ in the case of QFS. 
 In the lower parts of Figs.~\ref{fig2} and ~\ref{fig3} 
contributions to this effect due to 
 kinematics and dynamics are  shown. The five-fold differential 
cross section can be written as
\begin{eqnarray}
{d^5\sigma \over {d\Omega_1 d\Omega_2 dS}}  &=& ( \sum_{m_{in},m_{out}} 
 \vert <\phi_0 \vert U_0 \vert \phi > \vert^2  )~ \rho_{kin} ,
\label{eqz1}
\end{eqnarray}
with  the kinematical factor $\rho_{kin}$ containing the 
phase-space factor and 
 the initial flux. The transition probability for breakup 
$\vert <\phi_0 \vert U_0 \vert \phi > \vert^2$, 
averaged over the initial $m_{in}$ and summed 
over final $m_{out}$ sets of particles spin projections,  
 forms the
dynamical part of the cross section. In the lower parts of figures 
 the ratio of the relativistic 
to the nonrelativistic kinematical factor  
$\rho_{kin}^{rel} / \rho_{kin}^{nrel}$ as a function of S 
is shown  by the dashed 
line. The corresponding ratio for the dynamical parts of the cross section is 
shown by the solid line. 
 As seen in Fig.~\ref{fig3} for the QFS configuration 
 the whole 
  effect is due to a dynamical change of the transition matrix element. For 
this configuration the nonrelativistic and relativistic kinematical 
 factors 
are practically equal for large region of S-values (see Fig.~\ref{fig3}). 
For SSS about $30 \%$ of the total effect is due to a decrease of the 
relativistic kinematical factor with respect to the nonrelativistic 
one  (see Fig.~\ref{fig2}). 

The cross sections in these particular 
configurations are rather stable with respect 
to exchange between modern NN forces, combining them or not with three-nucleon 
forces~\cite{kuros}. Due to that 
 relativistic effects seem to explain the small and up to now puzzling  
 overestimation 
of the $65$~MeV SSS cros section data~\cite{psi} by modern nuclear forces 
and can account for the experimental width 
of this QFS peak~\cite{psiallet}.

\begin{figure}[hb]
\includegraphics[scale=0.59]{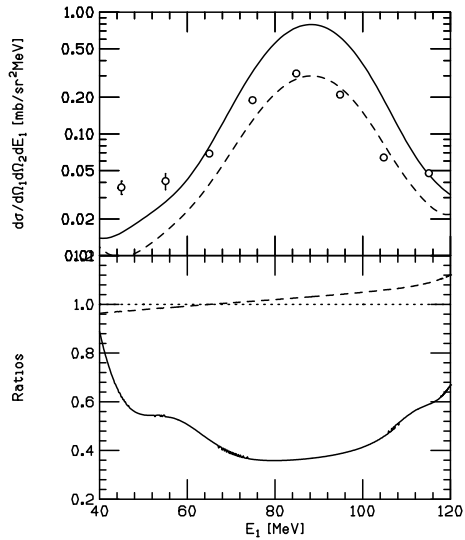}
\caption{
Five-fold differential cross section
 $d^5\sigma/d\Omega_1d\Omega_2dE_1$
 as a function of energy of the
 detected proton $E_1$
 for d(p,pn)p   reaction
 at $200$~ MeV. The polar lab. angle of the outgoing proton and neutron is
$\theta_1 = 52^{\circ}$ and  $\theta_2 = 45^{\circ}$, respectively.
The azimuthal angle
 $\phi_{12} = 180^{\circ}$.
For the description of lines and the bottom part see Fig.~\ref{fig2}. 
The  data are from \cite{br200}.
}
\label{fig4}
\end{figure}

At higher energies selectivity of breakup allows us to find the 
  configurations with significantly  larger  
 relativistic effects. 
 In Fig.~\ref{fig4} this is examplified at $E_{lab}^N = 200$~MeV 
and the predicted effects of up to $\approx 60 \%$, which are 
mostly of dynamical origin, seem to be supported by the data 
of ref.~\cite{br200}.

The selectivity of complete breakup is gradually lost 
 when uncomplete reactions 
are considered. In the total nd breakup cross section the effects disappear.  
 Integrating over all available 
complete breakup configurations provides nearly equal  relativistic 
(90.25 mb at 65 MeV and 43.37 mb at 250 MeV) and 
nonrelativistic (91.12 mb and 45.41 mb) total breakup cross sections.  
Also integrated elastic scattering angular distribution (71.25 mb 
and 9.33 mb - relativistic, and 71.40 mb and 9.57 mb - nonrelativistic) and 
 the total 
cross section for the nd interaction do not reveal significant 
relativistic effects.  This shows, 
that the discrepancies between theory and 
data found in previous studies 
 at higher energies  for the total cross 
section and elastic scattering angular distributions, 
 which  remain even after combining NN potentials with 3NFs, have to result 
from additional  contributions to the 3N force, which have different 
 than  the $2\pi$-exchange character. 

Summarizing, we showed that selectivity of the complete breakup reaction 
 enables 
 us to reveal  in 3N continuum clear signals from relativistic effects. 
Existing breakup data 
seem to support the predicted effects, when the relativity is 
included in the instant form of relativistic dynamics as proposed by Bakamjian 
and Thomas. Precise complete breakup data at energies around $200$~MeV are 
welcome to further test these predictions. The QFS breakup 
configurations due to their large cross sections and 
insensitivity to the details of nuclear forces 
 are favored for this purpose. 

This work was supported  by
the Polish Committee for Scientific Research under
Grant No.\ 2P03B00825 and by the Japan Society for the Promotion of Science.
HW would like to thank for hospitality and support during the
stay at RCNP, Japan and at TUNL, USA.
The numerical calculations have been performed on the
CRAY SV1 and on the IBM Regatta p690+  of the NIC in J\"ulich,
Germany.

\vspace{1.cm}

\bibliography{apssamp}

\end{document}